\documentclass[11pt,a4paper]{article}
\usepackage[hyperref]{naaclhlt2019}
\usepackage{times}
\usepackage{latexsym}
\usepackage{graphicx}
\usepackage{subcaption}
\usepackage{amsmath}
\usepackage{algorithm2e}

\usepackage{url}

\aclfinalcopy 

\title{A framework for streamlined statistical prediction using topic models}

\author{Vanessa Glenny$^{1,2}$ \\
  \hspace{25em}$^1$ARC Centre of Excellence for Mathematical and Statistical Frontiers (ACEMS)\\
  \hspace{26em}$^2$School of Mathematical Sciences, The University of Adelaide, SA 5005, Australia \\
  \hspace{7.7em}$^3$Data to Decisions CRC Stream Lead \\
  \hspace{11em}{\tt lewis.mitchell@adelaide.edu.au}
   \\\And
  Jonathan Tuke$^{1,2}$ \And
  Nigel Bean$^{1,2}$ \And
  Lewis Mitchell$^{1,2,3}$}

\date{}

\begin{document}
\maketitle

\begin{abstract}
In the Humanities and Social Sciences, there is increasing interest in approaches to information extraction, prediction, intelligent linkage, and dimension reduction applicable to large text corpora.
With approaches in these fields being grounded in traditional statistical techniques, the need arises for frameworks whereby advanced NLP techniques such as topic modelling may be incorporated within classical methodologies.
This paper provides a classical, supervised, statistical learning framework for prediction from text, using topic models as a data reduction method and the topics themselves as predictors, alongside typical statistical tools for predictive modelling. 
We apply this framework in a Social Sciences context (applied animal behaviour) as well as a Humanities context (narrative analysis) as examples of this framework.
The results show that topic regression models perform comparably to their much less efficient equivalents that use individual words as predictors.
\end{abstract}

\section{Introduction}

For the past 20 years, topic models have been used as a means of dimension reduction on text data, in order to ascertain underlying themes, or `topics', from documents. These probabilistic models have frequently been applied to machine learning problems, such as web spam filtering \cite{li2013}, database sorting \cite{krestel2009} and trend detection \cite{lau2012}. 

This paper develops a methodology for incorporating topic models into traditional statistical regression frameworks, such as those used in the Social Sciences and Humanities, to make predictions. 
Statistical regression is a supervised method, however it should be noted the majority of topic models are themselves unsupervised.

When using text data for prediction, we are often confronted with the problem of condensing the data into a manageable form, which still retains the necessary information contained in the text. 
Methods such as using individual words as predictors, or $n$-grams, while conceptually quite simple, have a tendency to be extremely computationally expensive (with tens of thousands of predictors in a model). 
Except on extremely large corpora, this inevitably leads to overfitting. 
As such, methods that allow text to be summarised by a handful of (semantically meaningful) predictors, like topic models, gives a means to use large amounts of text data more effectively within a supervised predictive context.

This paper outlines a statistical framework for predictive topic modelling in a regression context. First, we discuss the implementation of a relatively simple (and widely used) topic model, latent Dirichlet allocation (LDA) \cite{blei2003}, as a preprocessing step in a regression model. We then compare this model to an equivalent topic model that incorporates supervised learning, supervised LDA (sLDA) \cite{mcauliffe2008}.

Using topic models in a predictive framework necessitates estimating topic proportions for new documents, however retraining the LDA model to find these is computationally expensive. Hence we derive an efficient likelihood-based method for estimating topic proportions for previously unseen documents, without the need to retrain.

Given these two models hold the `bag of words' assumption (\textit{i.e.}, they assume independence between words in a document), we also investigate the effect of introducing language structure to the model through the hidden Markov topic model (HMTM) \cite{andrews2010}. The implementation of these three topic models as a dimension reduction step for a regression model provides a framework for the implementation of further topic models, dependent on the needs of the corpus and response in question.

\subsection{Definitions}

The following definitions are used when considering topic models.

\textit{Vocabulary }($V$): a set of $v$ unique elements (generally words) from which our text is composed.

\textit{Topic} ($\phi$): a probability distribution over the vocabulary. That is, for word $i$ in the vocabulary, a probability $p_{i} \in [0,1]$ is assigned of that word appearing, given the topic, with $\sum_{i = 1}^{v} p_{i} = 1$. In general, there are a fixed number $k$ of topics, $\boldsymbol{\phi} = \left\{\phi_{1},...,\phi_{k}\right\}$.

\textit{Document} ($\mathbf{w}$): a collection of $n_{j}$ units (or words) from the vocabulary. Depending on the topic model, the order of these words within the document may or may not matter.

\textit{Corpus} ($\mathbf{D}$): a collection of $m$ documents over which the topic model is applied. That is, $\mathbf{D} = \left\{\mathbf{w}_{1},...,\mathbf{w}_{m}\right\}$, each with length $n_{j}$, $j = 1,2,...,m$.

\textit{Topic proportion} ($\theta_{j}$): a distribution of topics over the document $j$. A corpus will then have an $m \times k$ matrix $\boldsymbol{\theta}$, where each row $j = 1,2,...,m$ corresponds to the distribution of topics over document $j$.

\section{LDA regression model}

Latent Dirichlet allocation (LDA) \cite{blei2003}, due to its simplicity and effectiveness, continues to be the basis for many topic models today. When considering topic regression, we take LDA as our `baseline' model; \textit{i.e.}, we measure all subsequent models against the performance of the LDA regression model.

LDA is an unsupervised process that assumes both topics and topic proportions are drawn from Dirichlet distributions. One reason for its simplicity is that it makes the `bag of words' assumption. LDA assumes the process outlined in Algorithm \ref{alg:lda} when generating documents.

\begin{algorithm}
	\For{$l = 1,2,...,k$}{
		generate the $k$ topics $\phi_{l} \sim \textrm{Dir}(\beta)$\;
	}
	\For{$j = 1,2,...,m$}{
		let $n_{j} \sim \textrm{Poisson}(\xi)$, the length of document $j$\;
		choose the topic proportions $\theta_{j} \sim \textrm{Dir}(\alpha)$\;
		\For{$i = 1,2,...,n_{j}$}{
			choose the topic assignment $z_{ji} \sim \textrm{Multi}(\theta_{j})$\;
			choose a word $w_{ji} \sim \textrm{Multi}(\phi_{z_{ji}})$\;
		}
		create the document $\boldmath{w}_{j} = \{w_{ji}\}_{i = 1,2,...,n_{j}}$\;
	}
\caption{LDA generative process.}
\label{alg:lda}
\end{algorithm}

Here, $\alpha$ (length $k$) and $\beta$ (length $v$) are hyperparameters of the distributions of the $\theta_{j}$ and $\phi_{l}$ respectively.

When topic modelling, we are generally interested in inferring topic proportions $\boldsymbol{\theta} = \left\{\theta_{1},...,\theta_{m}\right\}$ and topics $\boldsymbol{\phi}$ themselves, given the corpus $\mathbf{D}$. That is, we wish to find
\begin{eqnarray*}
	P\left(\boldsymbol{\theta},\boldsymbol{\phi} | \mathbf{D}, \alpha, \beta \right) = \frac{P\left(\boldsymbol{\theta},\boldsymbol{\phi}, \mathbf{D} | \alpha, \beta \right)}{P\left( \mathbf{D} | \alpha, \beta \right)}.
\end{eqnarray*}

The denominator, $P\left( \mathbf{D} | \alpha, \beta \right)$, the probability of the corpus, is understandably generally intractable to compute. For the purposes of this paper, we use collapsed Gibbs sampling as outlined in \citet{griffiths2004}, as an approximate method for finding the LDA model given the corpus.

\subsection{Regression model and number of topics}
\label{sec:lda_regression}

\begin{figure*}
	\centering
	\begin{subfigure}{.5\textwidth}
		\centering
		\includegraphics[width=\textwidth]{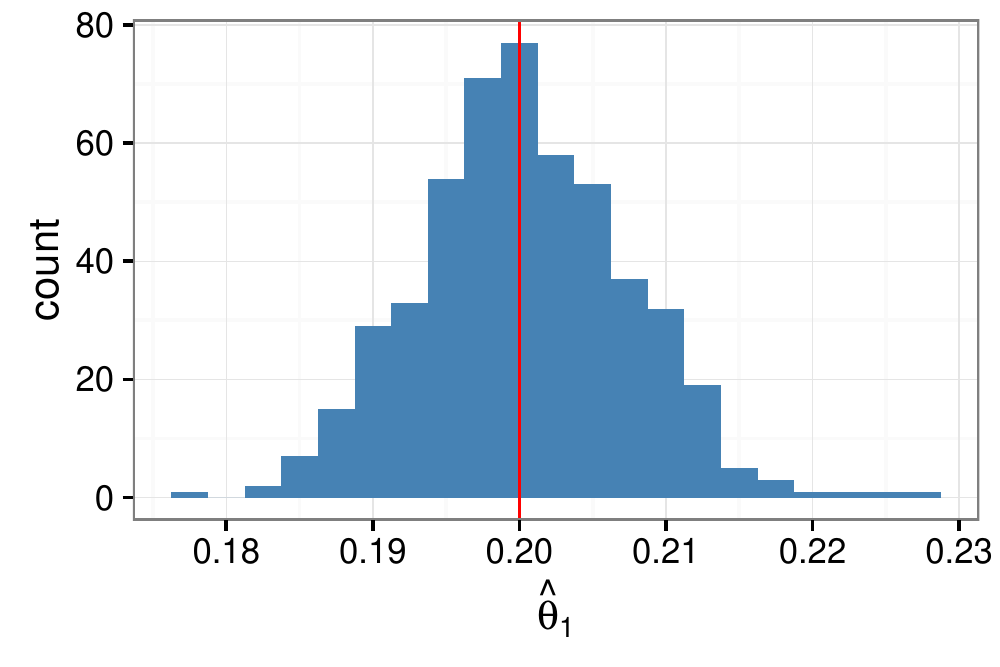}
		\caption{$\theta_{1}=0.2$}
		\label{fig:mle_2k_02}
	\end{subfigure}%
	\begin{subfigure}{.5\textwidth}
		\centering
		\includegraphics[width=\textwidth]{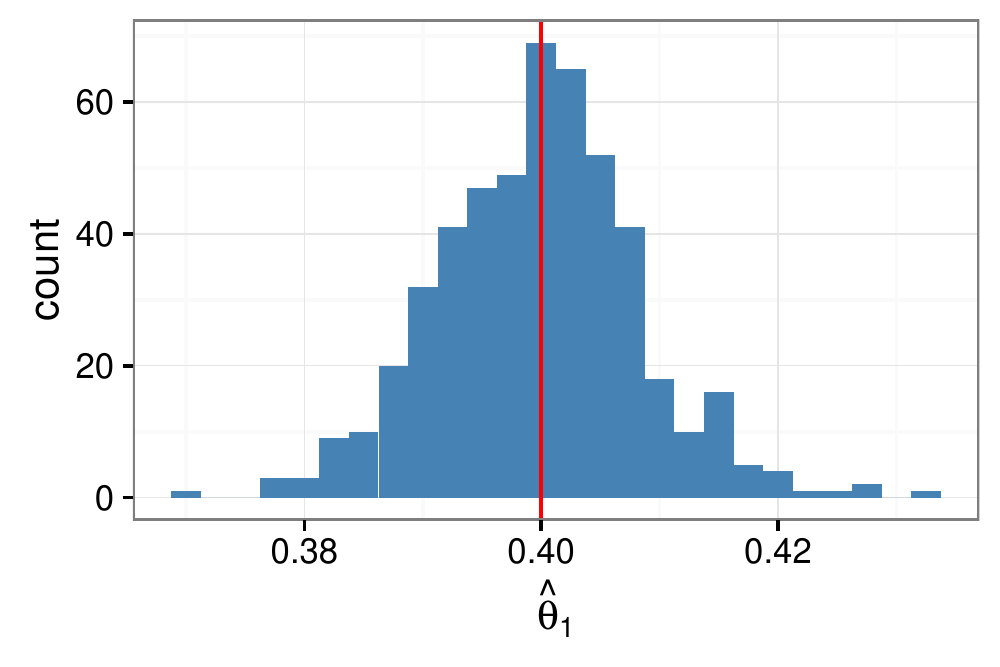}
		\caption{$\theta_{1}=0.4$}
		\label{fig:mle_2k_04}
	\end{subfigure}%
	\caption{Histograms of the maximum likelihood estimates of $\theta_{1}$ for corpora of two topics, given relative true values of $0.2$ and $0.4$.}
	\label{fig:mle_2k}
\end{figure*}

Given an LDA model on a corpus with some corresponding response variable, we use the topic proportions generated as predictors in a regression model. More specifically, we use the topic proportions $\boldsymbol{\theta}$ as the predictors, as the amount of a document belonging to each topic may be indicative of its response.

When applying LDA as a preprocessing step to a regression model, we must also bear in mind the number of topics $k$ we choose for the LDA model. While this number is assumed to be fixed in advance, there are various measures for determining the number that best `fits' the corpus, such as perplexity \cite{blei2003} and the log likelihood measure outlined in \citet{griffiths2004}.

However, given we are inferring this topic model with a specific purpose in mind, it would be prudent to include this information into the decision making process. For that reason, we choose the `best' number of topics $k$ to be the number that reduces the cross validation prediction error (CVPE) \cite{geisser1975} of the corresponding LDA regression model, found from $K$-fold cross validation of the model on the corpus. The CVPE is here defined to be
\begin{eqnarray*}
	\textrm{CVPE}_{K} = \sum\limits_{i = 1}^{K} \frac{m_{i}}{m} \textrm{MSE}_{i},
\end{eqnarray*}
where $K$ is the number of folds, $m_{i}$ is the number of documents in the $i$th fold, and $m$ the total number of documents in the corpus. The mean square error for the $i$th fold, denoted by $\textrm{MSE}_{i}$, is defined as
\begin{eqnarray*}
	\textrm{MSE}_{i} = \sum\limits_{j \in C_{i}} \frac{1}{m_{i}} \left( y_{j} - \hat{y}_{j} \right)^{2},
\end{eqnarray*}
where $\hat{y}_{j}$ is the model estimate of response $y_{j}$ for all documents in the set $C_{i}$, the $i$th fold. It follows that the better a model performs, the smaller the MSE and thus the CVPE.

While we choose the best number of topics based on the information in the regression model, it should be noted that LDA is still unsupervised, and that the topics have not been generated with the response in mind.

\subsection{Introducing new documents}

\begin{figure*}
	\centering
	\begin{subfigure}{.5\textwidth}
		\centering
		\includegraphics[width=\textwidth]{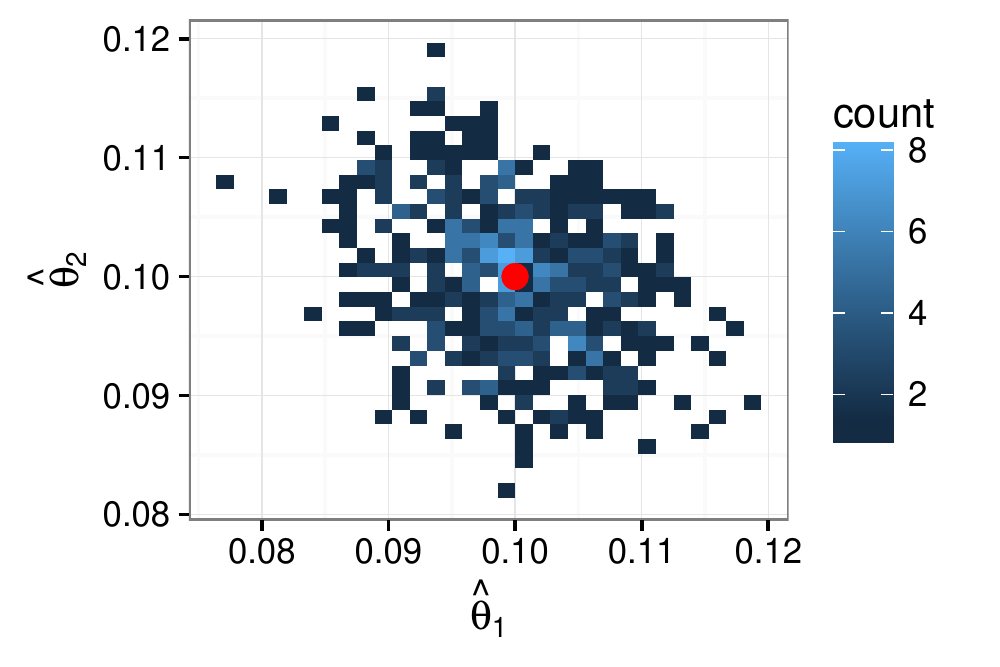}
		\caption{$\left\{\theta_{1},\theta_{2} \right\}=\left\{0.1,0.1\right\}$}
		\label{fig:mle_3k_01}
	\end{subfigure}%
	\begin{subfigure}{.5\textwidth}
		\centering
		\includegraphics[width=\textwidth]{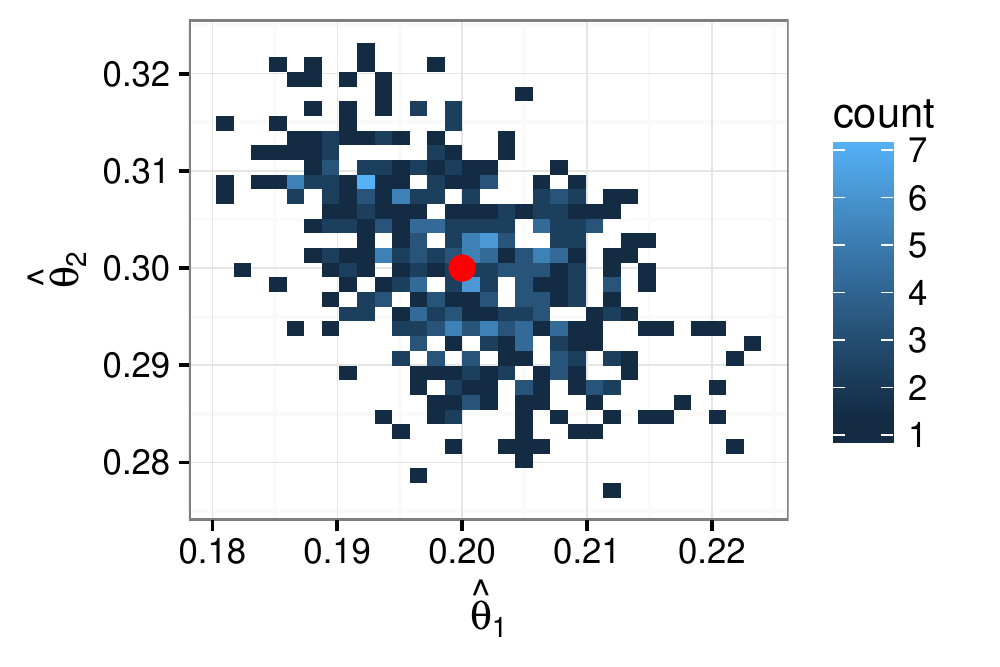}
		\caption{$\left\{\theta_{1},\theta_{2} \right\}=\left\{0.2,0.3\right\}$}
		\label{fig:mle_3k_02}
	\end{subfigure}%
	\caption{Histograms of the maximum likelihood estimates of $\left\{\theta_{1},\theta_{2} \right\}$ for corpora of three topics, given relative true values of $\left\{0.1,0.1\right\}$ and $\left\{0.2,0.3\right\}$.}
	\label{fig:mle_3k}
\end{figure*}

When it comes to prediction, we generally have a corpus for which we find our regression model, and use this model to predict the response of new documents that are not in the original corpus. Because our regression model requires us to know $\theta_{j}$, the topic proportion, for any new document $j$, we have two options. Either the topic model can be retrained with the new document added to the corpus, and the regression model retrained with the new topics on the old documents, or the topic proportions can be found based on the existing topic model.

For both efficiency's sake (\textit{i.e.}, to avoid retraining the model for every prediction), and for the sake of true prediction, the second option is preferable. Particularly in cross validation, it is necessary to have a completely distinct traning and test set of data. In retraining a topic model with new documents, we do not have a clear distinction between the two sets.

\citet{blei2003} outline a procedure for estimating the topic proportions of a held-out document, however this procedure follows a posterior approach that requires variationally inferring the posterior parameters, which are then used to approximate the expected number of words belonging to each topic, as an estimate for $\theta_{j}$.

We propose here a likelihood-based approach to estimation of topic proportions of new documents, by treating the problem as a case of maximum likelihood estimation. That is, we want to find $\hat{\theta}_{j}$, the estimate of $\theta_{j}$ that maximises the likelihood of document $j$ occurring, given our existing topic model. Therefore, we aim to maximise
\begin{eqnarray*}
	L(\theta_{j}) &=& f(\mathbf{w}_{j} | \theta_{j}) \\
    	&=& f(w_{j1},...,w_{jn_{j}} | \theta_{j}),
\end{eqnarray*}
where $w_{j1},...,w_{jn_{j}}$ are the words in document $j$. As LDA is a `bag of words' model, we are able to express this as
\begin{equation*}
	L(\theta_{j}) = \prod\limits_{i = 1}^{n_{j}} f(w_{ji} | \theta_{j}).
\end{equation*}
The law of total probability gives
\begin{equation*}
	L(\theta_{j}) = \prod\limits_{i = 1}^{n_{j}} \sum\limits_{l = 1}^{k} f(w_{ji} | z_{ji} = l, \theta_{j}) f(z_{ji} = l | \theta_{j}),
\end{equation*}
where $z_{ji}$ is the topic assignment for the $i$th word in document $j$. However, as the choice of word $w_{ji}$ is independent of the topic proportions $\theta_{j}$ given its topic assignment $z_{ji}$, we can write
\begin{equation*}
	L(\theta_{j}) = \prod\limits_{i = 1}^{n_{j}} \sum\limits_{l = 1}^{k} f(w_{ji} | z_{ji} = l) f(z_{ji} = l | \theta_{j}).
\end{equation*}
The likelihood is now expressed as the products of the topic proportions and the topics themselves.
\begin{eqnarray*}
	L(\theta_{j}) &=& \prod\limits_{i = 1}^{n_{j}} \sum\limits_{l = 1}^{k} \phi_{l,w_{ji}} \theta_{jl}.
\end{eqnarray*}
If we express the document as a set of word counts $\mathbf{N} = \{N_{1},...,N_{v}\}$, where $N_{i}$ is the number of times the $i$th word of the vocabulary appears in document $j$, then we can write the log likelihood of $\theta_{j}$ as
\begin{eqnarray*}
	l(\theta_{j}) = \mathbf{N} \cdot \log \left(\theta_{j} \boldsymbol{\phi} \right).
\end{eqnarray*}

In order to deal with words that appear in a new document, and not the original corpus, we assign a probability of 0 to any such word of appearing in any of the $k$ topics; this is equivalent to removing those words from the document.

To demonstrate the effectiveness of this method for estimation, we generate documents for which we know the topics and topic proportions. Suppose there exists a corpus comprising of two topics, with a vocabulary of 500 words. Given an assumed LDA model, we generate 500 documents with lengths between 5,000 and 10,000 words.

Given our newly generated documents, and known topics $\boldsymbol{\phi}$, we are able to test the validity of the MLE process outlined above by finding the estimates $\hat{\theta}_{j}$ for each document $j$ and comparing them to known topic proportions $\theta_{j}$. Figure \ref{fig:mle_2k} shows the results of the MLE method for finding topic proportion estimates for documents with certain true values of $\theta_{j}$. From these figures, there is a tight clustering around the true value $\theta_{j}$, and thus it is reasonable to assume that the MLE process for estimating the topic proportions of a new document given previously existing topics is sound. This process also holds for greater numbers of topics, as evidenced in Figure \ref{fig:mle_3k}, which estimates topic proportions for a three-topic document.

\section{sLDA regression model}

LDA is an unsupervised process, which does not take into account the response variable we are predicting when inferring topics. Several supervised methods have been developed to incorporate this knowledge, generally for the purpose of finding `better' topics for the corpus in question. Notably, supervised LDA (sLDA) \cite{mcauliffe2008} builds on the LDA model by assuming that some response $y_{j}$ is generated alongside each document $j = 1,2,...,m$ in the corpus, based on the topics prevalent in the document. When inferring the sLDA model, we are therefore inclined to find topics that best suit the response and therefore the prediction problem at hand.

Unlike LDA, we treat the topics $\boldsymbol{\phi}$ as unknown constants rather than random variables. That is, we are interested in maximising
\begin{eqnarray*}
	P\left( \boldsymbol{\theta}, \mathbf{z} | \mathbf{D}, \mathbf{y}, \boldsymbol{\phi}, \alpha, \eta, \sigma^{2} \right),
\end{eqnarray*}
where $\eta$ and $\sigma^{2}$ are parameters of the normally distributed response variable, $y_{j} \sim N(\eta^{T} \bar{z}_{j}, \sigma^{2})$, where $\bar{z}_{j} = (1/n_{j}) \sum_{i = 1}^{n_{j}} z_{ji}$.

As with LDA, this probability is computationally intractable, and thus we require an approximation method for model inference. For the purposes of this paper, we use a variational expectation-maximisation (EM) algorithm, as outlined in \citet{mcauliffe2008}.

When it comes to choosing the model with the most appropriate number of topics for the regression problem at hand, we use the same method as outlined for the LDA regression model in Section \ref{sec:lda_regression}.

The method behind sLDA is specifically developed for prediction. As such, we are able to compute the expected response $y_{j}$ from the document $\mathbf{w}_{j}$ and the model $\{ \alpha, \boldsymbol{\phi}, \eta, \sigma^{2}\}$. For a generalised linear model (as we use in this paper), this is approximated by
\begin{equation*}
	E\left[ Y_{j} | \mathbf{w}_{j}, \alpha, \boldsymbol{\phi},\eta, \sigma^{2} \right] \approx E_{q} \left[\mu\left(\eta^{T} \bar{\mathbf{z}}_{j} \right)\right],
\end{equation*}
where $\mu\left(\eta^{T} \bar{\mathbf{z}}_{j} \right) = E\left[Y_{j} | \zeta = \eta^{T} \bar{\mathbf{z}}_{j} \right]$ and $\zeta$ is the natural parameter of the distribution from which the response is taken. Again, further detail on this method is found in \citet{mcauliffe2008}.

\section{HMTM regression model}
\label{sec:hmtm}

Topic modelling is designed as a method of dimension reduction, and as such we often deal with large corpora that cannot otherwise be analysed computationally. Given the complexity of human language, we therefore have to choose what information about our corpus is used to develop the topic model. The previous two models, LDA and sLDA, have relied on the `bag of words' assumption in order to maintain computational efficiency. While for some corpora, the loss of all information relating to language and document structure may not have a particularly large effect on the predictive capability of the topic model, this may not hold for all prediction problems.

One simple way of introducing structure into the model is through a hidden Markov model (HMM) structure \cite{baum1967,baum1970}; in fact, there already exist multiple topic models which do so. We look here at the hidden Markov topic model (HMTM) \cite{andrews2010}, which assumes that the topic assignment of a word in a document is dependent on the topic assignment of the word before it. That is, the topic assignments function as the latent states of the HMM, with words in the document being the observations. The HMTM assumes the generative process outlined in Algorithm \ref{alg:hmtm} for documents in a corpus.
\begin{algorithm}[h]
 	\For{$l = 1,2,...,k$}{
  		generate topics $\phi_{l} \sim \textrm{Dir}(\beta)$\;
	}
	\For{$j = 1,2,...m$}{
		generate starting probabilities $\boldsymbol{\pi}_{j} \sim \textrm{Dir}(\alpha)$\;
		\For{$l = 1,2,...,k$}{
			generate the $l$th row of the transition matrix, $\boldsymbol{\Theta}_{j}$, $\Theta_{jl} \sim \textrm{Dir}(\gamma_{l})$\;
		}
		choose the topic assignment for the first word $z_{j1} \sim \textrm{Multi}(\boldsymbol{\pi}_{j})$\;
		select a word from the vocabulary $w_{j1} \sim \textrm{Multi}(\phi_{z_{j1}})$\;
		\For{$i = 2,3,...,n_{j}$}{
			choose the topic assignment $z_{ji}$ based on transition matrix $\boldsymbol{\Theta}_{j}$\;
			select a word from the vocabulary $w_{ji} \sim \textrm{Multi}(\phi_{z_{ji}})$\;
		}
		create the document $\mathbf{w}_{j} = \left\{ w_{ji} \right\}_{i = 1,...,n_{j}}$\;
	}
\caption{HMTM generative process.}
\label{alg:hmtm}
\end{algorithm}

Here, $\alpha$, $\beta$ and $\boldsymbol{\gamma} = \left\{ \gamma_{1},...,\gamma_{k} \right\}$ are Dirichlet priors of the starting probabilities, topics and transition probabilities respectively.

When it comes to prediction, we are able to use the transition matrices for each document $\boldsymbol{\Theta}_{j}$ as predictors, but to keep consistency with the previous models we take the equilibrium distributions of the matrices as the topic proportions $\theta_{j}$. That is, we find $\theta_{j}$ such that
\begin{eqnarray*}
	\theta_{j} \boldsymbol{\Theta_{j}} = \theta_{j}, \quad \textrm{and} \quad \theta_{j} \mathbf{e} = 1.
\end{eqnarray*}
This also fits with the concept of topic models as a form of dimension reduction, allowing $k-1$ variables, as opposed to $k(k-1)$ when using the transition matrix $\boldsymbol{\Theta}_{j}$. As models are often fit using hundreds of topics \cite{blei2012, griffiths2004}, this makes models faster to compute. We choose the number of topics $k$ here with the same method outlined in Section \ref{sec:lda_regression}.

\subsection{Introducing new documents}

Like with the LDA regression model, we require a method for estimating the topic proportion $\theta_{j}$ of any new documents from which we are predicting a response, that does not involve retraining the entire model. To do so, we rely on techniques used for HMMs; specifically, we use a modified Baum-Welch algorithm.

The Baum-Welch algorithm is used as an approximate method to find an HMM $\Omega = \{ \boldsymbol{\Theta}, \boldsymbol{\phi}, \boldsymbol{\pi} \}$, given some observed sequence (in this case, a document). However, the key difference here is that our emission probabilities (or topics) $\boldsymbol{\phi}$ are common across all documents in our corpus, and thus when introducing any new documents for prediction we assume that we already know them. Given the Baum-Welch algorithm calculates forward and backward probabilities based on an assumed model, and updates estimates iteratively, we may simply take our assumed $\boldsymbol{\phi}$ found from the initial HMTM as the truth and refrain from updating the emission probabilities.

We are generally dealing with very small probabilities in topic modelling - $\boldsymbol{\phi}$ generally has tens of thousands of columns (the length of the vocabulary) over which probabilities must sum to one. While in theory this does not change how we would approach parameter estimation, computationally these probabilities are frequently recognised as zero. To make the process more numerically stable, we implement the adapted Baum-Welch algorithm demonstrated and justified in \citet{shen2008}.

While we are ultimately interested in finding topic proportions $\theta_{j}$ for prediction, the Baum-Welch algorithm finds the transition matrix $\boldsymbol{\Theta_{j}}$ for some document. We are able to deal with this in the same way as finding the original HMTM regression model, by taking $\theta_{j}$ to be the equilibrium probabilities of $\boldsymbol{\Theta_{j}}$.

\section{Testing the topic regression models}

To demonstrate the use of topic models in a regression framework, we apply them to a problem involving online advertisements. Specifically, we have a corpus containing 4,151 advertisements taken from the trading website, Gumtree\footnote{\url{www.gumtree.com.au}}, pertaining to the sale of cats in Australia, and hand-labelled by an expert. 
Of these advertisements, 2,187 correspond to relinquished cats and 1,964 to non-relinquished.
We train a model to predict `relinquished status' from the text of an advertisement, using a topic regression model. A cat is considered to be relinquished if it is being given up by its owner after a period of time, as opposed to cats that are sold, either by breeders or former owners.

In order to improve efficiency and model quality, we first clean our text data. Details on the cleaning steps can be found in Appendix \ref{sec:app_clean}.

\subsection{Word count model}
Before investigating regression models that use topic proportions as predictors, it is worth developing a `gold standard' model, \textit{i.e.}, a model whose predictive capability we aim to match with our topic regression models. Because the problem here involves a relatively small corpus (advertisements with a median word count of 35), we are able to compare our topic regression models to a model that uses individual words as its predictors.

In a much larger corpus, this kind of prediction would be cumbersome to compute - hence our reliance on topic models and other dimension reduction techniques.

Because we are predicting a categorical, binary variable, we use logistic regression. Rather than using all words in the corpus (as this would drastically overfit the model), we use a step-up algorithm based on the Akaike information criterion (AIC) \cite{akaike1974} to choose the most significant words for the model, without overfitting.

Instead of applying the step-up process to the entire vocabulary (of exactly 13,000 words), we apply it to the 214 most common words (\textit{i.e.}, words that appear in at least 2.5\% of the documents in the corpus). The chosen model uses 97 predictors, with coefficients appearing consistent with what you would expect from the problem: for example, the word \textit{kitten} is indicative of non-relinquished advertisements, while \textit{cat} is the opposite, which is expected as younger cats are less likely to be relinquished.

To assess the predictive capability of this and other models, we require some method by which we can compare the models. For that purpose, we use receiver operating characteristic (ROC) curves as a visual representation of predictive effectiveness. ROC curves compare the true positive rate (TPR) and false positive rate (FPR) of a model's predictions at different threshold levels. The area under the curve (AUC) (between 0 and 1) is a numerical measure, where the higher the AUC is, the better the model performs.

\begin{figure}
	\centering
	\includegraphics[width=0.48\textwidth]{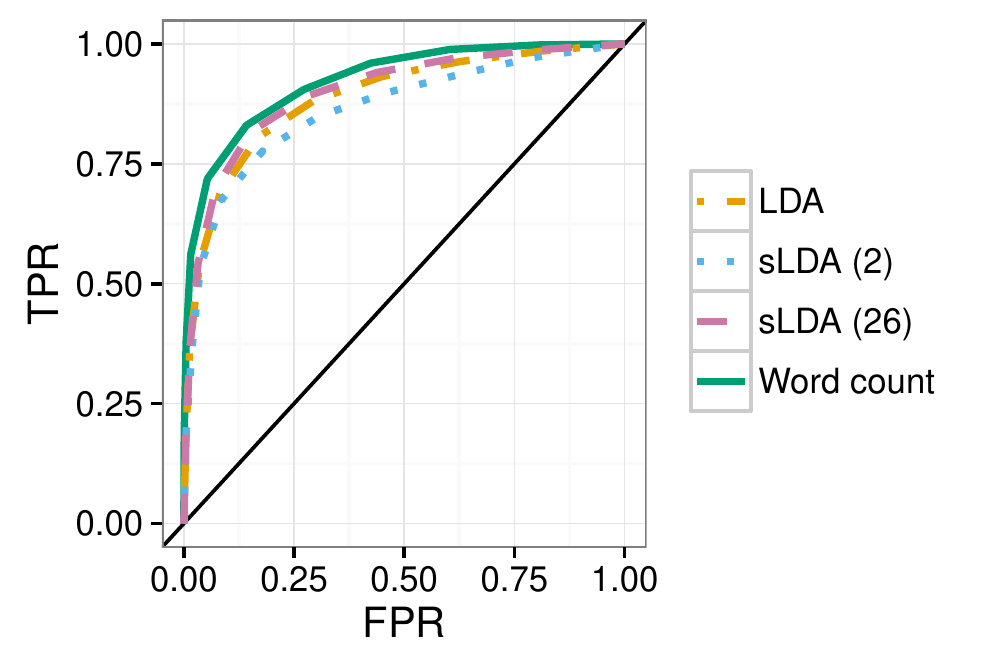}
	\caption{Threshold-averaged ROC curves of the word count model, LDA regression model, and sLDA regression models with two and 26 topics respectively.}
	\label{fig:gum_roc_comp}
\end{figure}

We cross-validate our model by first randomly splitting the corpus into a training set (95\% of the corpus) and test set (5\% of the corpus). We then fit the model to the training set, and use it to predict the response of the documents in the test set. We repeat this process 100 times. The threshold-averaged ROC curve \cite{fawcett2006} is found from these predictions, and shown in Figure \ref{fig:gum_roc_comp}. 
Table \ref{tab:gum_auc} shows the AUC for each model considered.

\begin{table}[h!]
\centering
\begin{tabular}{lrr }
	\hline
	\textbf{Model} & \textbf{AUC} & \textbf{95\% CI} \\
	\hline
	Word count & 0.9264 & $(0.9234, 0.9294)$\\
	LDA & 0.8913 & $(0.8871, 0.8955)$\\
	sLDA (2 topics) & 0.8588 & $(0.8534, 0.8642)$ \\
	sLDA (26 topics) & 0.9030 & $(0.8988, 0.9073)$ \\
	\hline
	
\end{tabular}
\caption{TArea under the curve (AUC) for the models used on the Gumtree dataset, with their 95\% confidence intervals.}
\label{tab:gum_auc}
\end{table}

\subsection{Topic regression models}
Using the method outlined in Section \ref{sec:lda_regression}, we choose the LDA regression model with 26 topics as the `best' for this problem. 
Inspection of the top words included in these 26 topics shows individual topics associated with different breeds (e.g., `persian', `manx') as well as urgency of selling (e.g., `urgent', `asap'), suggesting that the model is not overfit to the data. 
We generate a threshold-averaged ROC curve using the same cross validation method as earlier, yielding an area under the curve (AUC) of $0.8913$. The curve can be seen in Figure \ref{fig:gum_roc_comp}. While not as high as the AUC for the word count model, the LDA regression model is significantly more efficient, taking only $3\%$ of the time to calculate.

We can compare this result to that of an sLDA regression model. The model chosen for this problem has two topics, giving a threshold-averaged ROC curve under cross validation with an AUC of $0.8588$. It is surprising that the LDA regression model should outperform sLDA, as sLDA incorporates the response variable when finding the most appropriate topics. However, this can be attributed to the number of topics in the model: the sLDA regression model with 26 topics outperforms the LDA model, with an AUC of $0.9030$.

The word count model still outperforms the sLDA model, however once again the topic regression model is significantly more efficient, taking only $0.6\%$ of the time to calculate. Further details on the models and their calculation can be found in Appendix \ref{sec:app_tm}.

\section{Incorporating language structure}

When evaluating the usefulness of incorporating document structure into a topic model for regression, we require a corpus and problem that we would expect would be heavily influenced by this structure. To understand the predictive capability of the HMTM regression model over that of the more simplistic LDA, we therefore consider predicting the storylines of the 2003 film \textit{Love Actually}\footnote{\url{www.imdb.com/title/tt0314331/}}, known for its interwoven yet still quite distinct storylines. We therefore ask if we are able to predict to which storyline a scene belongs, based on the dialogue in that scene.

The film consists of 79 scenes, each pertaining to one of 10 storylines. The scenes were hand-classified by storyline, and their dialogue forms the documents of our corpus. We once again clean our data; more detail can be found in Appendix \ref{sec:app_clean}.

\subsection{Word count model}
As with the Gumtree dataset, we first construct a word count model against which we can measure the performance of our topic regression models. Once again, this can be done because we are working with a small corpus; otherwise, we would generally consider this approach to be computationally too heavy.

As we have a categorical, non-binary response variable (storyline) with 10 levels, we use a multinomial logistic regression model. We again use a step-up process with AIC as the measure to determine which words in our vocabulary to use as predictors in our model. As our vocabulary consists of only 1,607 unique words, we consider all of them in our step-up process. After applying this process, the model with three predictors, \textit{minister}, \textit{night} and \textit{around}, is chosen.

We are no longer able to easily apply ROC curves as a measure of performance to this problem, as we are dealing with a non-binary response. We instead use a Brier score \cite{brier1951}, a measure for comparing the predictive performance of models with categorical responses. The Brier score is
\begin{eqnarray*}
	\textrm{BS} = \frac{1}{m} \sum\limits_{j=1}^{m} \sum\limits_{i=1}^{s} \left( \hat{y}_{ji} - o_{ji} \right)^{2},
\end{eqnarray*}
where $\hat{y}_{ji}$ is the probability of document $j$ belonging to storyline $i$, and $o_{ji} = 1$ if document $j$ belongs to storyline $i$, and $0$ otherwise, for document $j = 1,2,...,m$ and storyline $i = 1,2,...,s$. Each term in the sum goes to zero the closer the model gets to perfect prediction, and as such our aim is to minimise the Brier score in choosing a model.

For each document in the corpus, we find the probabilities of each outcome by using the remaining 78 documents (or training dataset) as the corpus in a multinomial logistic regression model with the same three predictors as found above. Due to the fact that the training dataset here is smaller than the Gumtree dataset, we perform leave-one-out cross validation on each document in the corpus (rather than using a 95/5 split). We then predict the outcome based on the words found in the left-out document (or test dataset), and repeat for all 79 scenes. However, due to the short length of some scenes, and the fact that unique words must be thrown out, we restrict the testing to 57 of the 79 scenes: the remaining scenes do not generate a numerically stable approximation for $\theta_{j}$ for the HMTM regression model.

The Brier score calculated using this method for the step-up word count model is $0.8255$.

\subsection{Topic regression models}
For the LDA regression model for this problem, we determine the `best' number of topics $k$ to be 16. As with the word count model, we use the Brier score to evaluate the performance of this model compared to others in the chapter. We again use the leave-one-out cross validation approach to predict the probabilities of a scene belonging to each storyline.

The Brier score found for the LDA regression model is $1.6351$. While this is higher and therefore worse than the Brier score for the word count model above, this is not unexpected and we are more interested in seeing how the LDA model fares against other topic models.

We compare these results to the HMTM regression model, as outlined in Section \ref{sec:hmtm}. We choose the model with 12 topics, according to the CVPE. The Brier score calculated from 57 scenes for the HMTM regression model is $1.5749$. While still not up to the standard of the word count model at $0.8255$, this appears to be a slight improvement on the LDA model, meaning that dropping the `bag of words' assumption may in fact improve the predictive performance of the model. However, it should be kept in mind that the LDA model is better at handling short documents. It would be worth applying these models to corpora with longer documents in future, to see how they compare. Further details on the computation of these models can be found in Appendix \ref{sec:app_tm}.

One of the motivating ideas behind having topic dependencies between consecutive words, as in the HMTM model, is that some documents will have a predisposition to stay in the same topic for a long sequence, such as a sentence or a paragraph. This argument particularly applies to narrative-driven corpora such as the \textit{Love Actually} corpus. To that end, we may adapt the HMTM described above so that the model favours long sequences of the same topic, by adjusting the Dirichlet priors of the transition probabilities, $\boldsymbol{\gamma} = \{ \gamma_{1},...,\gamma_{k} \}$, to favour on-diagonal elements. By specifying these priors to be
\begin{equation*}
		\gamma_{ls} = \begin{cases}
			0.99 + 0.01/k \quad \text{if} \quad l = s\\
			0.01/k \quad \text{elsewhere},
	\end{cases}
\end{equation*}
for $l = 1,2,...,k$, we choose the persistent HMTM regression model with three topics. This results in a Brier score of $0.9124$, which is a massive improvement on the original HMTM regression model and makes it very competitive with the word count model. 
Table \ref{tab:love_hard_class} summarises these results.

\begin{table}[ht]
\centering
\begin{tabular}{lrr}
	\hline
	\textbf{Model} & \textbf{Accuracy} & \textbf{Brier score}\\
	\hline
	Word count & 26.58 & 0.8255\\
	LDA & 12.66 & 1.6351\\
	HMTM & 14.04 & 1.5749\\
	Persistent HMTM & 15.58 & 0.9124\\
	\hline
\end{tabular}
\caption{Table of the percentage of hard classifications of storylines for each left-out scene in the corpus that are correct, alongside the Brier score, for each model.}
\label{tab:love_hard_class}
\end{table}

\section{Discussion and further research}
This paper outlines and implements a streamlined, statistical framework for prediction using topic models as a data processing step in a regression model. In doing so, we investigate how various topic model features affect how well the topic regression model makes predictions.

While this methodology has been applied to three specific topic models, the use of any particular topic model depends heavily on the kind of corpus and problem at hand. For that reason, it may be worth applying this methodology to incorporate different topic models in future, depending on the needs of the problem at hand.

In particular, we investigate here the influence of both supervised methods, and the incorporation of document structure. A logical next step would be to propose a model that incorporates these two qualities, in order to see if this improves predictive capability on corpora with necessary language structure.

\bibliography{bib_masters}
\bibliographystyle{acl_natbib}

\appendix

\section{Appendix}
\label{sec:supplemental}

\subsection{Text cleaning}
\label{sec:app_clean}

The following steps were taken to clean the Gumtree corpus:
\begin{itemize}
	\item removal of punctuation and numbers,
	\item conversion to lower case,
	\item removal of stop words (\textit{i.e.}, common words such as \textit{the} and \textit{for} that contribute little lexically), and
	\item removal of grammatical information from words (\textit{i.e.}, stemming).
\end{itemize}

When stemming words in this paper, we use the stemming algorithm developed by Porter for the Snowball stemmer project \cite{porter2001}. Similarly, when removing stop words, we use the (English language) list compiled, again, in the Snowball stemmer project.

In cleaning the \textit{Love Actually} corpus, we perform the first three steps outlined here. However, unlike with the Gumtree dataset, we do not stem words, as grammatical information is more pertinent when incorporating language structure.

\subsection{Topic model inference}
\label{sec:app_tm}

For each topic model, we choose the best number of topics from models generated with between two and 40 topics.

For the LDA models found in this paper, we use the \textit{LDA} function from the R package \textbf{topicmodels}, with the following parameters:
\begin{itemize}
	\item $\tt{burnin} = 1000$,
	\item $\tt{iterations} = 1000$, and
	\item $\tt{keep} = 50$.
\end{itemize}

The sLDA model in this paper was found using the $\tt{slda.em}$ function from the R package \textbf{lda}, with the following parameters:
\begin{itemize}
	\item $\tt{alpha} = 1.0$,
	\item $\tt{eta} = 0.1$,
	\item $\tt{variance} = 0.25$,
	\item $\tt{num.e.iterations} = 10$, and
	\item $\tt{num.m.iterations} = 4$.
\end{itemize}

We use the Python code from \citet{andrews2010} for the generation of our HMTM.

\end{document}